\documentclass[11pt,twoside]{article}
\usepackage{asp2010}
\resetcounters

\begin{document}
\label{authorguide}

\allowtitlefootnote
\title{Empirical Calibration of the Physical Parameters of Solar-like stars using $uvby$-Str\"omgren Photometry}
\author{G.~Dalle Mese,$^1$ O.~L\'opez-Cruz,$^1$ W. J.~Schuster,$^2$, C. Chavarr{\'\i}a-K$^2$, J. G. Garc\'ia$^{2 \dagger}$}
\affil{$^1$ Instituto Nacional de Astrof\'isica, \'Optica y Electr\'onica (INAOE), Luis Enrique Erro 1, C.P. 72840, Tonantzintla,
Puebla, M\'exico\\
$^2$Instituto de Astronom{\'\i}a, UNAM. Apartado Postal 877, Ensenada, B.C., C.P. 22800, M\'exico}

\begin{abstract}
We present an empirical calibration of the physical parameters of A, F and early G-type stars of luminosity class V. We have used a statistical approach based on a sample of about 15 000 stars having both $uvby$-Str\"omgren photomerty \citep{hauck_1998} and spectral types taken from SIMBAD. Stars closer than 70 pc have been considered reddening-free. In this paper we present the results for 1900 stars within 70 pc. Mean unreddened measurements have been used as input to CHORIZOS \citep{maiz_2004}. By assuming ``solar metallicity" ([Fe/H]=0), we have been able to determine effective temperatures ($T_{eff}$) and surface gravities ($\log g$) for each spectral type. We have found a tight correlations among the observations and the derived physical parameters; for example, $T_{eff}$ can be expressed as a cubic polynomial in $\left(b-y\right)_0$. From the distribution of colors and indices for 128 stars and their associated physical parameters we have proposed an alternative definition for solar-analogs: stars whose classification is G2 V with $\left(b-y\right)_0 =0.3927\pm0.0190$, $\left(m_1\right)_0 = 0.1901\pm0.0143$, and $\left(c_1\right)_0=0.3302 \pm0.0388$,  $T_{eff}=5930 \pm145$, and $\log g =4.70 \pm 0.33$. These values are consistent with previous empirical results; however, $\left(m_1\right)_0$ differs from a recent model-based value for this index derived by \cite{melendez}. This discrepancy points out the actual solar metallicity problem. We suggest that our sample of solar analogs could be useful for searches of solar twins and exoplanets. 
\end{abstract}

\section{Introduction}
\label{introduction}
Intermediate-band photometry has clear advantages over broad-band photometry: its results are closer to those of low-resolution spectroscopy and allow the separation of stellar physical parameters such as temperature, surface gravity and metallicity. Specifically, the $uvby$-Str\"omgren system \citep{strom_1966} was carefully defined to closely match and reproduce stellar spectral types. This system considers the following linear combinations of indices: $c_1=(u-v)-(v-b)$, and $m_1=(v-b)-(b-y)$, which measure the Balmer Jump and line blanketing for A-G type stars, respectively. As the stellar physical properties change with spectral type, $c_{1}$ is a good temperature indicator for O to A-type stars, but becomes a luminosity indicator for A to F-type stars. A similar behavior is exhibit by the $m_{1}$ index working as a metallicity or peculiarity indicator around A-type stars, but as a chemical composition indicator for F to G-type stars. The color ($b-y$) is sensitive to temperature but not metallicity (because of comparable and small line blanketing in $b$ and $y$) and the ($v-b$) color is very sensitive to metallicity for F and G stars due to larger blanketing in $v$.

\subsection{The MK System of Spectral Classification}

The MK is an autonomous and self consistent system of spectral classification. It relies on a well-defined set of standard stars, chosen on the basis of the phenomenology of spectral lines, blends, and bands, according to a general progression of color index (abscissa) and luminosity (ordinate) \citep{morgan}. The standard stars define an array that is located on the two-dimensional spectral type vs. luminosity class diagram.

Below we show that the $uvby$-Str\"omgren photometric system does reproduce MK spectral types. We found that  the distribution of colors and indices for each spectral type permits the definition of photometric classification boxes. MK types can be assign unambiguously within two spectral subclasses. We have derive the physical parameters using mean color and indices by spectral type. We have found a new calibration for $T_{eff}$ and $\left(b-y\right)_0$. These results have have allowed us to provide an alternative definition for solar analogs. 

\section{Observational Data}
The compilation of Str\"omgren photometric measurements from \cite{hauck_1998}, that contains $\sim$ 63 000 stars was cross-matched  with the \textit{Set of Identifications, Measurements, and Bibliography for Astronomical Data} (SIMBAD), the HIPPARCOS database \citep{perry} to produce a list of stars with photometric colors and indices, spectral types, and parallaxes. Our final sample with reliable spectral types (A0 V to G5 V) and photometric measurements ($V \lesssim 18^m.0 $ for G2 V stars) contains $ \sim $15 000 stars. 

\section{Interstellar Reddening}
Understanding the local interstellar reddening is of fundamental importance in the derivation of the $\left(b-y\right)_0$ calibration, and likewise for stellar metallicities, distance and ages. The nearest interstellar dust patches are at about 70 pc in some particular directions, this region devoid of dust is commonly identified with the \textit{Local Hot Bubble}. Indeed, several investigations  indicate that  the effects of interstellar reddening are almost negligible for stars closer than 70 pc from the Sun \citep[e.g.,][]{tim, leroy, holm,luck}. In this work we present the analysis for 1900 stars  within 70 pc from the Sun; since no correction for extinction is needed their colors and indices are taken as they are (i.e., $(b-y) = \left(b-y\right)_0, m_1 = \left(m_1\right)_0$, and $c_1 = \left(c_1\right)_0$. The rest of the stars were corrected for galactic extinction as described in \cite{crawford_1975} and \cite{schuster_1989}. A full analysis on the total sample of 15 000 stars will be presented elsewhere.

\section{The Distribution of Observed Properties} 
\begin{figure}
\centering
\includegraphics[angle=90, scale=0.24]{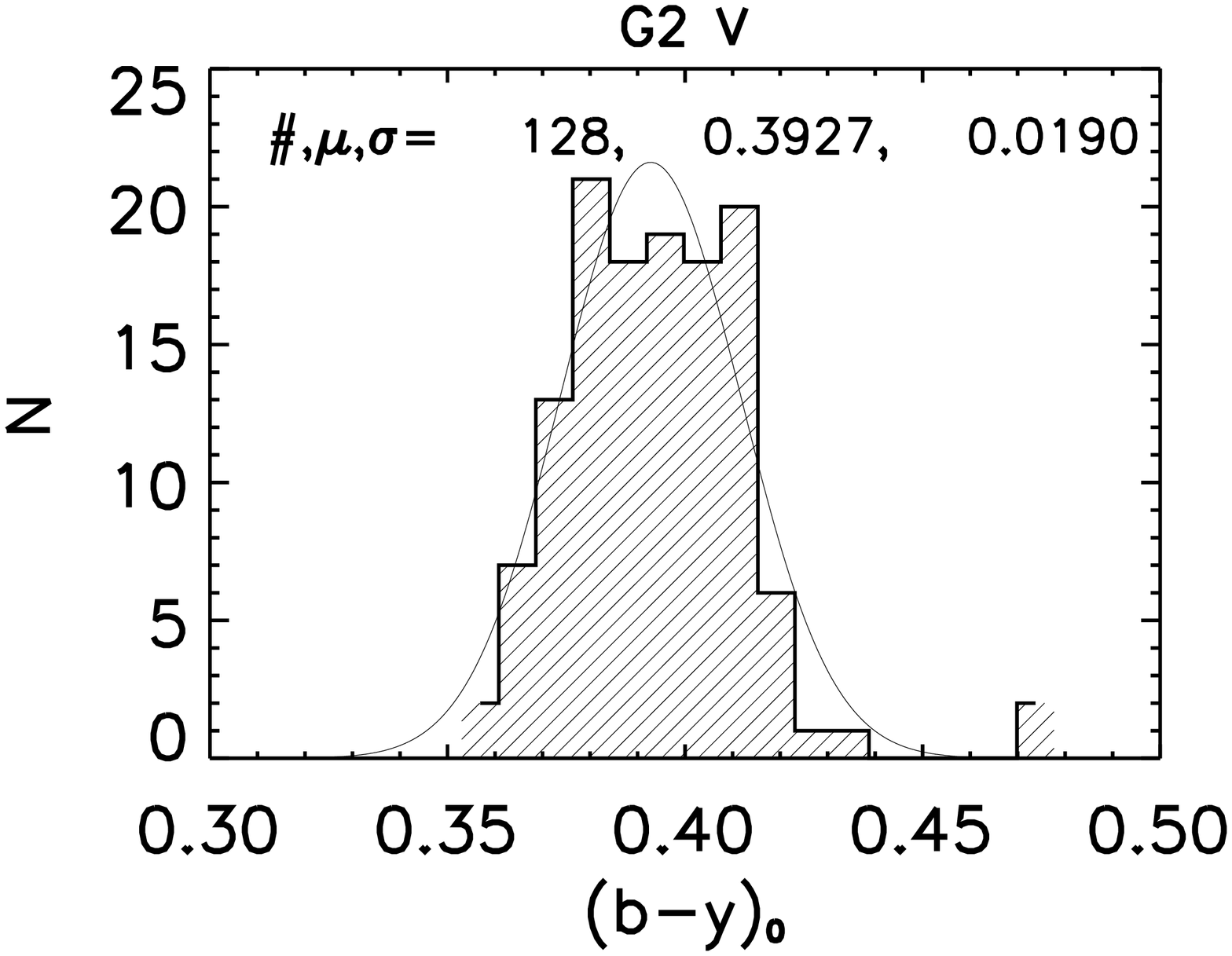}
\includegraphics[angle=90, scale=0.24]{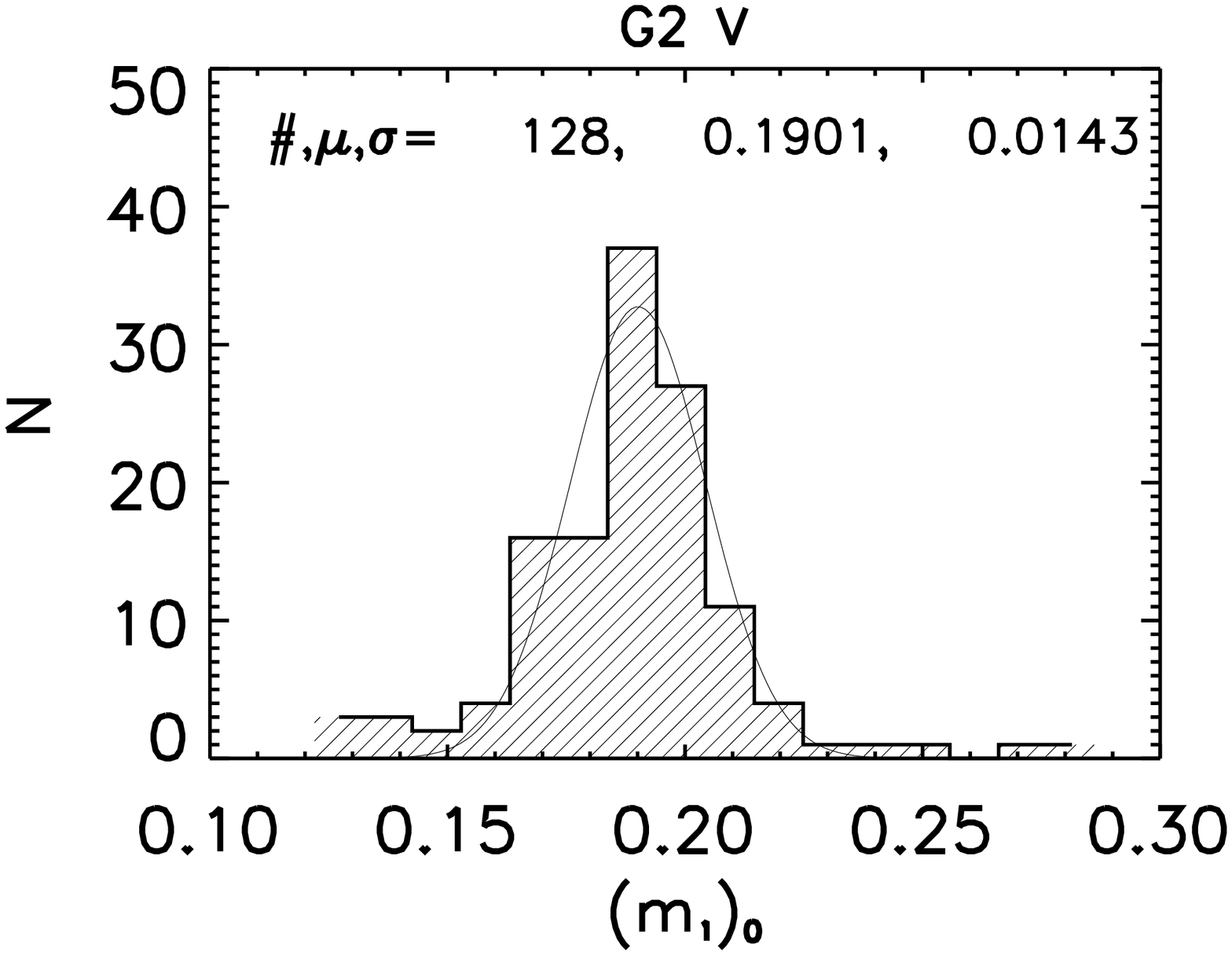}
\includegraphics[angle=90, scale=0.24]{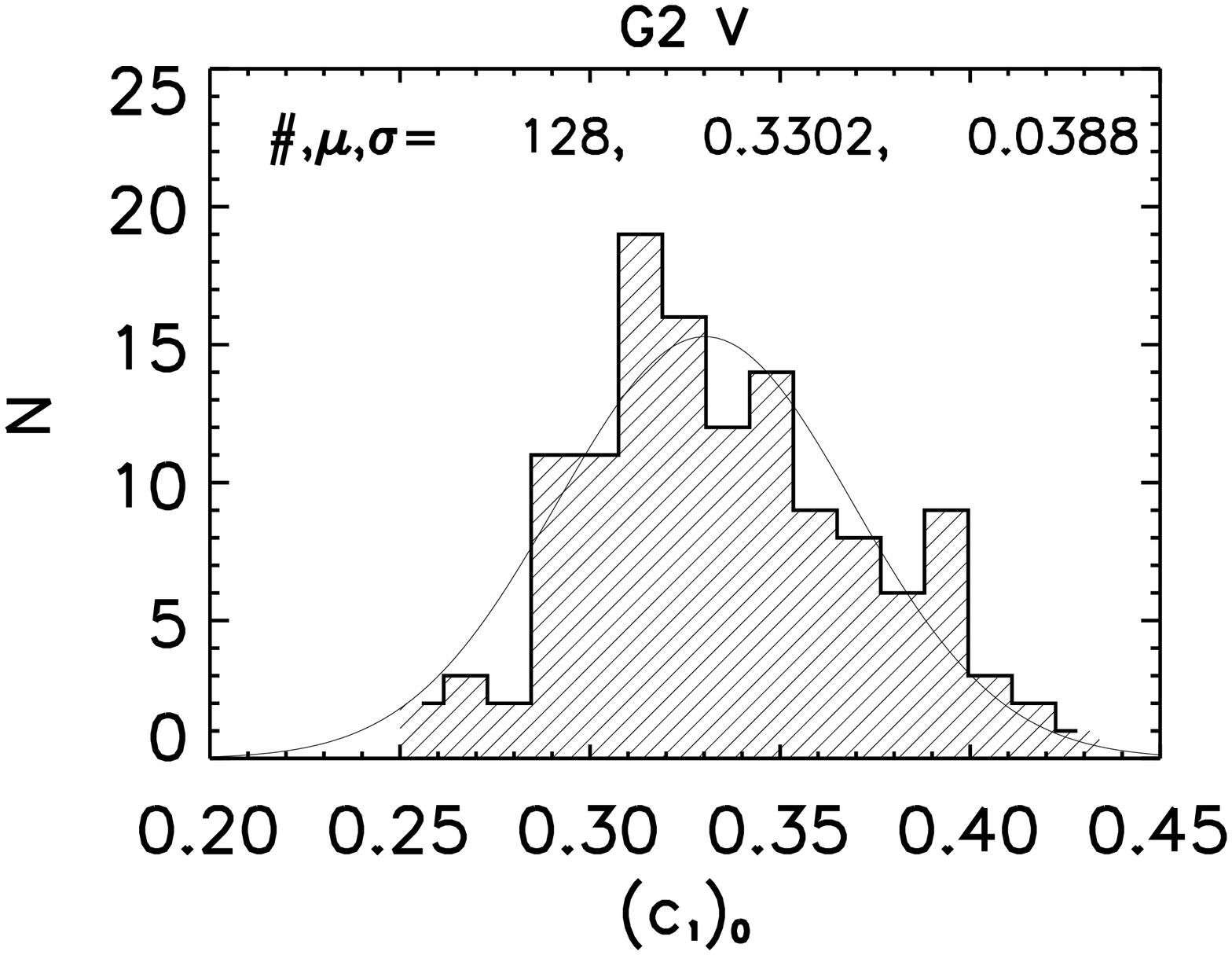}
\caption{The panels show the statistical distribution of the Str\"omgren photometric indices for nearby G2 V stars (distances $<$70 pc). The mean values of these distributions are: $\left(b-y\right)_0 =0.3927\pm0.0190$, $\left(m_1\right)_0=0.1901\pm0.0143$ and $\left(c_1\right)_0=0.3302 \pm0.0388$. With these values we can provide an alternative definition for  solar analogs, see the text for details.}
\label{histos}
\end{figure}

We have generated the distributions of colors and color indices by spectral type for main sequence stars from A0 to G5. In Figure \ref{histos} we show that the distribution of colors and indices for G2 V stars closely follow Gaussian distributions. This is expected given the discreetness of the spectral types, the small photometric errors reported (of the order of a thousand of a magnitude), the size of the sample considered (up to 200 stars per spectral type), and the De Moivre-Laplace Theorem. Highly discordant values will occupy the tails of the distribution while most of the values will cluster around the mean, as the number of stars gets larger the distribution becomes narrower. We should note that the entrances in the \citet{hauck_1998} catalog are weighted values, while the spectral types were taken from different sources in the literature.  

The mean values and widths ($\sigma$) of the distribution of  measured color and indices can be read directly from  Table 1. The distributions of spectral types shows a paucity at A5 V to A6 V, this was pointed out by \citet{houk}, we have neglected A6 V stars because only fours stars were found in \cite{hauck_1998} within 70 pc. If we consider these mean values and their widths for the distribution of $\left(b-y\right)_0$ by spectral type; then, we can classify stars with a resolution of two spectral classes for stars of luminosity class V. We have checked the Str\"omgren color and indices for the MK standard stars in the grid defined by \citet{LC91}. We found that the published values fall within $1 \sigma$ from the mean values for the corresponding spectral type. Hence, we suggest that our statistical approach provides an accurate  way to determine MK types photometrically. 

\begin{figure}
\centering
\includegraphics[scale=0.45,angle=90]{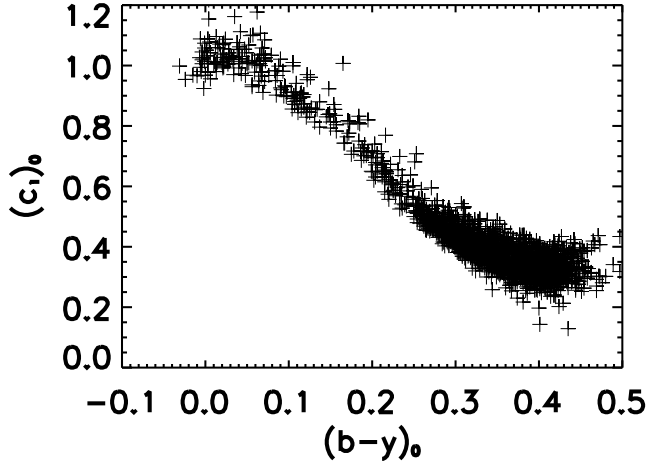}
\caption{The locus of A0-G5 dwarfs in the $\left(c_1\right)_0$, $\left(b-y\right)_0$ plane. $\left(c_1\right)_0$ measures the Balmer Jump (surface gravity) and $\left(b-y\right)_0$ the temperature. We expect to find a maximum at $\left(b-y\right)_0$ $\sim 0$ and, if observed values of $\left(b-y\right)_0$ are used, some scatter toward larger values of this index is due to interstellar reddening.}
\label{balmer}
\end{figure}

Figure \ref{balmer} shows the distribution for the 1900 A0-G5 V stars in the $\left(c_1\right)_0$ vs. $\left(b-y\right)_0$ diagram. The broadening in $\left(c_1\right)_0$ is produced by surface gravity \citep[e.g.,][]{Lester86}; hence, in this diagram we can appreciate the sensitivity of the spectral classification to surface gravity. As expected, for main sequence stars, the Balmer Jump reaches its maximum around A0 V stars. For earlier than A0 stars the $\left(m_1\right)_0$ is no longer a surface gravity indicator, it becomes a temperature indicator, instead. 
 
 The results presented here are in agreement and complement the results of \cite{crawford84}. Table 1 is in close agreement, but improves over the results of \citet{Oblak} due to better sampling for some spectral types and better handle on the reddening correction.

\begin{table}\scriptsize
 \begin{center}
\begin{center}
  \caption{\scriptsize Average Photometric Str\"omgren Colors and Indices by MK spectral types}
\end{center}

 \begin{tabular}{lcccccccc}

  \hline
 \hline

  \multicolumn{1}{c}{\tiny Spectral} &
   \multicolumn{1}{c}{\tiny $\left(b-y\right)_0$}&
  \multicolumn{1}{c}{\tiny $\sigma \left(b-y\right)_0$}&
  \multicolumn{1}{c}{\tiny $\left(m_1\right)_0$} &
  \multicolumn{1}{c}{\tiny $\sigma \left(m_1\right)_0$ }&
  \multicolumn{1}{c}{\tiny $\left(c_1\right)_0$}&
  \multicolumn{1}{c}{\tiny $\sigma \left(c_1\right)_0$}&
  \multicolumn{1}{c}{\tiny $T_{eff}$}&
  \multicolumn{1}{c}{\tiny $\log g$}\\
\tiny Type & & & & & & & \tiny (K)&\tiny $cm/s^2$\\
\hline

A0 V &	0.0004 &  0.0140  &	0.1609 & 0.0256 & 1.0317 &  0.0646 & 	9949 $\pm$ 405 & 4.41 $\pm$ 0.39\\ 
A1 V &	0.0177 &  0.0205  &	0.1786 & 0.0140 & 1.0043 &  0.0431 & 	9672 $\pm$ 383 & 4.41 $\pm$ 0.31\\ 
A2 V &	0.0306 &  0.0261  &	0.1802 & 0.0222 & 1.0241 &  0.0333 & 	9445 $\pm$ 442 & 4.37 $\pm$  0.32\\ 
A3 V &	0.0559 &  0.0186  &	0.1835 & 0.0150 & 1.0114 &  0.0837 & 	8907 $\pm$ 420 & 4.52 $\pm$  0.36\\ 
A4 V &	0.0705 &  0.0131  &	0.1928 & 0.0119 & 0.9432 &  0.0739 & 	8647 $\pm$ 273 & 4.54 $\pm$  0.32\\ 
A5 V &	0.0957 &  0.0301  &	0.1919 & 0.0162 & 0.9514 &  0.0331 & 	8458 $\pm$ 506 & 4.33 $\pm$  0.30\\ 
A7 V &	0.1136 &  0.0116  &	0.1997 & 0.0147 & 0.9054 &  0.0351 & 	8109 $\pm$ 259 & 4.11 $\pm$  0.28\\ 
A8 V &	0.1502 &  0.0115  &	0.1825 & 0.0081 & 0.8326 &  0.0856 & 	7747 $\pm$ 206 & 4.34 $\pm$  0.37\\ 
A9 V &	0.1781 &  0.0488  &	0.1717 & 0.0148 & 0.7605 &  0.1067 & 	7623 $\pm$ 569 & 4.63 $\pm$  0.35\\ 
F0 V &	0.1980 &  0.0387  &	0.1690 & 0.0236 & 0.7133 &  0.1048 & 	7423 $\pm$ 414 & 4.67 $\pm$  0.36\\ 
F1 V &	0.2221 &  0.0552  &	0.1619 & 0.0121 & 0.6144 &  0.0897 & 	7261 $\pm$ 538 & 4.66 $\pm$  0.33\\ 
F2 V &	0.2551 &  0.0340  &	0.1531 & 0.0137 & 0.5412 &  0.0793 & 	7000 $\pm$ 325 & 4.59 $\pm$  0.37\\ 
F3 V &	0.2715 &  0.0147  &	0.1492 & 0.0112 & 0.4800 &  0.0385 & 	6874 $\pm$ 148 & 4.39 $\pm$  0.37\\ 
F4 V &	0.2899 &  0.0218  &	0.1563 & 0.0252 & 0.4644 &  0.0466 & 	6713 $\pm$ 199 & 4.50 $\pm$  0.40\\ 
F5 V &	0.2988 &  0.0153  &	0.1538 & 0.0103 & 0.4324 &  0.0431 & 	6649 $\pm$ 145 & 4.45 $\pm$  0.40\\ 
F6 V &	0.3196 &  0.0166  &	0.1546 & 0.0123 & 0.4149 &  0.0523 & 	6478 $\pm$ 145 & 4.53 $\pm$  0.40\\ 
F7 V &	0.3345 &  0.0165  &	0.1587 & 0.0134 & 0.3881 &  0.0432 & 	6362 $\pm$ 143 & 4.56 $\pm$  0.39\\ 
F8 V &	0.3499 &  0.0152  &	0.1678 & 0.0132 & 0.3755 &  0.0472 & 	6244 $\pm$ 121 & 4.60 $\pm$  0.38\\ 
F9 V &	0.3600 &  0.0164  &	0.1687 & 0.0199 & 0.3477 &  0.0480 & 	6171 $\pm$ 137 & 4.66 $\pm$  0.35\\ 
G0 V &	0.3733 &  0.0172  &	0.1757 & 0.0136 & 0.3429 &  0.0413 & 	6068 $\pm$ 134 & 4.67 $\pm$  0.34\\ 
G1 V &	0.3832 &  0.0161  &	0.1862 & 0.0220 & 0.3417 &  0.0394 & 	6001 $\pm$ 118 & 4.68 $\pm$  0.34\\ 
G2 V &	0.3927 &  0.0190  &	0.1901 & 0.0143 & 0.3302 &  0.0388 & 	5930 $\pm$ 145 & 4.70 $\pm$  0.33\\ 
G3 V &	0.4042 &  0.0171  &	0.1979 & 0.0179 & 0.3336 &  0.0417 & 	5847 $\pm$ 133 & 4.71 $\pm$  0.32\\ 
G4 V &	0.4130 &  0.0153  &	0.2099 & 0.0186 & 0.3284 &  0.0385 & 	5787 $\pm$ 108 & 4.73 $\pm$  0.31\\ 
G5 V &	0.4253 &  0.0175  &	0.2220 & 0.0229 & 0.3168 &  0.0347 & 	5715 $\pm$ 125 & 4.74 $\pm$ 0.30\\ 
\hline	
\label{tab}
\end{tabular}
\end{center}
\end{table}

\section{Physical Parameters}
\label{physicalparameters}
We have used CHORIZOS \citep{maiz_2004} to obtain $T_{eff}$ and $\log g$. This code compares photometric data of a star with model spectral-energy distributions. The code calculates the likelihood for the full specified parameter ranges (extinction, temperature, gravity, metallicity), thus allowing for the identification of multiple solutions and the evaluation of the full correlation matrix for the derived parameters of a single solution.

The input parameters that we used were the mean $uvby$-Str\"omgren colors of each spectral type with their respective uncertainties and a family of spectral-energy distributions (SED) of Kurucz with the following parameters: $T_{eff}=[3 500,50 000]K$, $\log g=[0.0,5.0$], $\log (Z/Z_{\odot})=[-1.5,0.0]$, $E(B-V)= [-0.5, 5.0]$ and the total to selective extinction ratio for the interstellar medium of $R_{V}=3.1$. CHORIZOS performs a $\chi^2$ analysis to assign a likelihood to each of the models in a $N$ dimensional space by comparing the observed and the set of modeled colors. In our case, $\log (Z/Z_{\odot})=0$, $E(B-V)= 0$ where held fixed. The results are presented in Table 1, columns 8 and 9. The errors are given by CHORIZOS, they are meant to represent the range of values for each spectral type. 

\section{Results}
\label{results}

The main results from this work are presented in Table 1. We can see a clear progression between $\left(b-y\right)_0$ and $T_{eff}$, a simple fit to third order polynomial accounts for temperature variations in the range between A0 V and G5 V stars. The fit is given by the following expression (see Figure 3):
\begin{eqnarray}
T_{eff}&=&9955.49(\pm 38.79) -20010.40(\pm 843.26)\left(b-y\right)_0 \nonumber \\ 
&& + 45881.00(\pm 4589.29)\left(b-y\right)_0^2-53299.20(\pm 6944.62)\left(b-y\right)_0^3.
\end{eqnarray}

These results should be taken with the caveat that our statistical approach overlooks for the effects of rotation, magnetic fields, metallicity and that the derived physical parameters come from unidimensional model atmospheres in LTE. However, we have found a nice correspondence among spectral types, $\left(b-y\right)_0$, and $T_{eff}$. Equation 1 is in agreement with previous calibrations \citep[e.g.,][]{alonso, gray_2001, clem_2004}. However, we find that our result for the index $\left(m_1\right)_0$ is inconsistent with the solar value calculated by \cite{melendez}. We may explain this discrepancy in the apparent problem with the solar metallicity that has pervaded in the literature and was discussed during this conference. The newest 3D model atmospheres have changed the abundance value for the Sun \citep[e.g.,][]{asplund}.

\subsection{An Alternative Definition for Solar Analogs}

The Sun has never been classified using a regular telescope and spectrograph. It always appears resolved and the  effects of limb darkening, rotation, variability creep in hampering the classification\footnote{Prof. Robert F. Garrison used to start his lectures of G stars by stating that the worse classified star in the MK system was the Sun. He taught a graduate course on stellar classification for many years at the University of Toronto.}. Alternative standards for the G2 V type have been considered, for example, HR483, Jupiter IV and 18 Sco, among others. Depending on adopted set of standards, apparent differences in the photometric colors may arise for each spectral type (E. Mamajek, private communication). 

There is a lively discussion regarding the definition of a solar twin, that is a star with mass, chemical composition, age, $T_{eff}$,  $\log g$, luminosity, magnetic filds, velocity fields, equatorial rotation, etc. very similar if not identical to those of the Sun \citep{cayrel}. However, it has been argued that no solar twin has yet been found \citep{strass} and the search seems to be far from over \citep[e.g.,][]{grayc}. Nevertheless, the general consensus is that 18 Sco is the best solar twin candidate. 
 
We wish to introduce an alternative definition from that of \cite{cayrel} for solar analogs. We propose that G2 V stars with $\left(b-y\right)_0 = 0.3927 \pm 0.0190$, $\left(m_1\right)_0= 0.1901 \pm 0143$, and $\left(c_1\right)_0 = 0.3302 \pm 0.0388 $ whose $T_{eff} = 5930 \pm 145$ and $\log g = 4.70 \pm  0.33$ should be considered solar analogs. This could help to narrow down the searches for solar twins. Solar twins are an important piece in the understanding of the origin of our solar system.

\begin{figure}
\centering
\includegraphics[scale=0.45,angle=90]{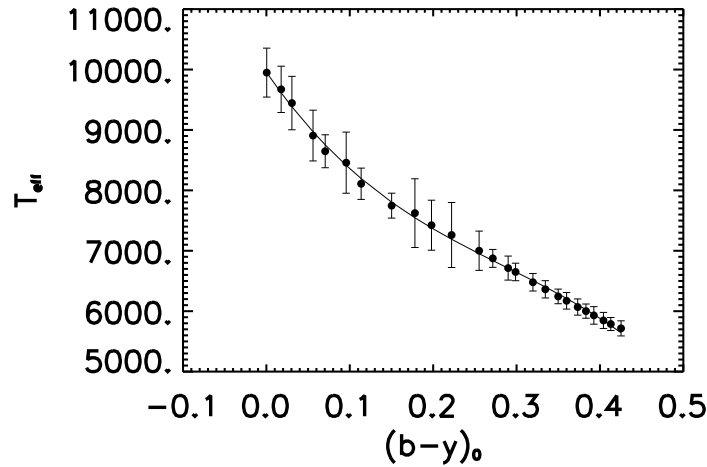}
\caption{Relation between the derived $T_{eff}$ and $\left(b-y\right)_0$ index for A0 V to G5 V stars of Table 1, each point represents an spectral type, errors in the calibrated temperature are given by CHORIZOS  \citep[cf.,][]{clem_2004}, the solid line is a cubic  polynomial fit: $T_{eff}=9955.49(\pm 38.79) -20010.40(\pm 843.26)\left(b-y\right)_0 + 45881.00(\pm 4589.29)\left(b-y\right)_0^2-53299.20(\pm 6944.62)\left(b-y\right)_0^3$.}
\label{teff}
\end{figure}

\section{Concluding Remarks}
\begin{itemize}

\item We have shown the effectiveness of the Str\"omgren system in the investigation on the physical properties of stars, avoiding the degeneracies inherent to broad band photometry. We will apply this technique to stellar clusters and galaxies.

\item We have introduced a technique that allows the derivation of accurate $T_{eff}$ for main sequence stars from a comparison of mean reddening-free photometric indices for $\sim$1900 stars with a grid of synthetic spectra and fluxes computed from Kurucz models. The $T_{eff}$ can be expressed as a cubic polynomial in $\left(b-y\right)_0$. Our calibration is in agreement with previous ones \citep[e.g.,][]{alonso, gray_2001, clem_2004}.

\item From the distribution of colors and indices for 128 stars and their associated physical parameters we have proposed an alternative definition for solar-analogs: stars whose classification is G2 V with $\left(b-y\right)_0=0.3927\pm0.0190$, $\left(m_1\right)_0=0.1901\pm0.0143$, and $\left(c_1\right)_0=0.3302\pm0.0388$, $T_{eff}=5930\pm145$ and $\log g=4.70\pm0.33$.

\end{itemize}

\acknowledgements{ We wish to thank Eric Mamajek and Miguel Ch\'avez for enlightening discussions. We are grateful to Laura Parrao for checking some of the results presented here and for reading the manuscript. We acknowledge  financial support from Direcci\'on de Formaci\'on Acad\'emica and Coordinaci\'on de Astrof{\'\i}sica of INAOE. We are also thankful the organizers of this very fruitful and enjoyable meeting.}

\bibliography{dallemese_pv}

\end{document}